\documentclass{aip-cp}
\usepackage[numbers]{natbib}
\usepackage{rotating}
\usepackage{graphicx}

\begin{document}

\title{Symmetry Analysis In Inflationary Cosmology}
\author[aff1]{Andronikos Paliathanasis\corref{cor1}}
\eaddress[url]{anpaliat@phys.uoa.gr}
\affil[aff1]{Institute of Systems Science, Durban University of Technology, PO Box 1334,
Durban 4000, RSA} \corresp[cor1]{Corresponding author: anpaliat@phys.uoa.gr}
\maketitle

\begin{abstract}
We approach the cosmological inflation though symmetries of differential
equations. We consider the general inflaton field in a homogeneous
Friedmann--Lema\^{\i}tre--Robertson--Walker spacetime and with the use of
conformal transformations we are able to write the generic algebraic
solution for the field equations. We put emphasis on the inflationary models
and we show how we can construct new inflationary models from already known
models by using symmetry transformations.
\end{abstract}



\section{INTRODUCTION}

In the late 19th century Sophus Lie had the idea to bring to the
differential equations the algebraic success of infinitesimal
transformations which had attended the question of the solution of
polynomial equations. While his initial expectations were never realized, he
was able to establishe a new method for the determination of solutions for
differential equations. His pioneering work was published in 1888 \cite%
{lie1,lie2,lie3} with the title \textquotedblleft Theory of transformation
groups\textquotedblright\ and defined a new area in mathematics, the
Symmetries. The concept of Lie symmetries has been one of the main materials
for the study of nonlinear differential equations, the main results which
establish the Lie theory a main mathematical tool in modern science can be
found in the works of Ovsiannikov \cite{Ovsi}, Ibragimov \cite{ibra}, Olver
\cite{olver}, Crampin~\cite{Crampin}, Leach \cite{leach000} and many others.

There is a wide range of applications of the theory of symmetries in natural
sciences and specifically in physics, from analytical mechanics \cite{sym1},
to particle physics \cite{sym2}, and gravitational physics \cite{sym5}. An
important class of symmetries which are commonly used in General Relativity
are the spacetime collineations \cite{sym5}. Collineations are the
generators of continuous transformations where geometric objects of the
gravitational theory are transformed under a specific rule, or remain
invariant. The most important class of collineations are the isometries, or
Killing vectors (KV), which leaves invariant the metric tensor of the
theory. Furthermore, the existence of collineations can simplify the
Einstein-field equations and provide important kinematic or dynamical
constraints in the physical properties of the theory \cite{con2}. A
classification of exact solutions in General Relativity according to the
admitted collineations of the spacetime can be found in \cite{con3}.

In General Relativity, the gravitational field equations is a set of ten
nonlinear partial differential equations. Collineations of the Einstein
tensor pass through the field equations as Lie symmetries of differential
equations and in particular reduce the number of the independent variables,
or the number of the independent equations. However, these Lie symmetries
are restricted only to the space of independent variables, while in general
there can exist continuous transformations in the space of the dependent
variables which leave the field equations invariant. Hence, a detailed study
of the symmetries of the field equations is required. Indeed such an
analysis was the main subject of study in \cite{mahleach,wafo}.

The cosmological principle, which states that the spatial distribution in
the universe is homogeneous and isotropic in large scales, is supported by
the detailed analysis of the recent cosmological data \cite{data1,data2}.
Mathematically, the cosmological principle is expressed with the existence
of a six isometries (KVs) in the physical spacetime, which means that the
only possible line element which describes the universe in large scales is
the Friedmann--Lema\^{\i}tre--Robertson--Walker (FLRW) metric
\begin{equation}
ds^{2}=-N^{2}\left( t\right) dt^{2}+a^{2}\left( t\right) \left( \frac{dr^{2}%
}{1-kr^{2}}+r^{2}\left( d\theta^{2}+\sin^{2}\theta d\phi ^{2}\right) \right)
\label{me.01}
\end{equation}
in which $a\left( t\right) $ is the scale factor of the universe, $N\left(
t\right) $ is the lapse function where without loss of generality can be set
a nonzero constant, and $k$ denotes the spatial curvature of the three
dimensional space, while the cosmological data indicates that $k\simeq0$.

A theoretical mechanics which explains the large-scale structure of the
universe is the inflation \cite{guth}. Inflation is attributed to the
existence of the \textquotedblleft inflaton\textquotedblright\ which drives
a period of acceleration in the early universe \cite{Aref1}. More
specifically the inflaton, plays the role of a matter source in Einstein's
GR with negative pressure which displays an antigravity behavior \cite{lid02}%
. The inflaton it is assumed that is described by a scalar field. The
introduction of the scalar field in Einstein's GR introduces new degrees of
freedom in the theory which can drive the dynamics such that the
inflationary era to be described.\ The physical particle representation of
the inflaton scalar field is still unknown it can describe an
\textquotedblleft exotic\textquotedblright\ nonvisible matter source or the
degrees of freedom which follow from modifications of Einstein-Hilbert's
Action \cite{Aref1}.

The field equations in Einstein' GR are of second-order, however in the
presence of the scalar field, the field equations show unexpected
complexity, as a result, exact and analytic solutions exist only for few
scalar field potentials, see \cite{sfr01,sfr04,sfr05,sfr06,sfr07,sfr08} and
references therein.

The application of the Lie's theory in the gravitational field equations and
more specifically in cosmological problems have been proved very useful and
new exact and analytic solutions have been determined in scalar field/scalar
tensor theories and in modified/alternative theories of gravity \cite%
{ns01,ns03,ns04,ns05}, for a recent review we refer the reader in \cite%
{revsym}. However, besides the construction of analytic solutions, the
symmetries of differential equations have been used also as a method of
quantization in quantum cosmology \cite{qs01,qs02}. In this work, we focus
on the application of the theory for the symmetries of differential
equations in inflationary models and more specifically on the theoretical
description of the inflaton field. The plan of the paper is as follows.

In Section \ref{sec2} we present the cosmological model of our consideration
as also the main set of differential equations. Moreover, the main
mathematical tools of our analysis are discussed. In Section \ref{sec3} we
present a family of inflationary models which follow from a set of maximally
symmetric master equations. We apply Lie theory to construct new
inflationary models. Our conclusions are presented in Section \ref{sec4}.

\section{SCALAR FIELD COSMOLOGY}

\label{sec2}

In the context of GR and in the presence of a scalar field $\phi\left(
x^{c}\right) $ the Action Integral of the gravitational field equations is
\cite{scot1}%
\begin{equation}
S=\int dx^{4}\sqrt{-g}R+\int dx^{4}\sqrt{-g}\left( \frac{1}{2}g_{ab}\phi
^{;a}\phi^{;b}-V\left( \phi\right) \right) ,   \label{me.02}
\end{equation}
where $R$ is the Ricciscalar of the underlying manifold with metric $g_{ab}$%
, and $V\left( \phi\right) $ is the potential which drives the dynamics of
the field $\phi\left( x^{k}\right) $.

Variation with respect to the metric tensor of (\ref{me.02}) provides the
Einstein field equations%
\begin{equation}
R_{ab}-\frac{1}{2}Rg_{ab}=\phi_{;a}\phi_{;b}-\frac{1}{2}g_{ab}\left(
\phi_{;c}\phi^{;c}+V\left( \phi\right) \right) ,   \label{me.03}
\end{equation}
while variation with respect to the field $\phi\left( x^{i}\right) $ gives
the \textquotedblleft Klein-Gordon\textquotedblright\ equation $\phi
_{;ab}g^{ab}+V_{,\phi}=0.$

For the spatially flat FLRW spacetime the Ricciscalar is calculated to be $R=%
\frac{6}{N^{2}}\left( \frac{\ddot{a}}{a}+\left( \frac{\dot{a}}{a}\right)
^{2}-\frac{\dot{a}\dot{N}}{aN}\right) ,~$while by assuming that the scalar
field inherits the symmetries of the spacetime it follows $\phi\left(
x^{i}\right) =\phi\left( t\right) $, consequently, the latter equations are
written%
\begin{equation}
3H^{2}-\frac{1}{2}\dot{\phi}^{2}-V\left( \phi\right) =0,   \label{me.06}
\end{equation}%
\begin{equation}
2\dot{H}+3H^{2}+\left( \frac{1}{2}\dot{\phi}^{2}-V\left( \phi\right) \right)
=0,   \label{me.07}
\end{equation}
and%
\begin{equation}
\ddot{\phi}+3H\dot{\phi}+V_{,\phi}=0,   \label{me.08}
\end{equation}
in which $H\left( t\right) $ is the Hubble constant defined as $H\left(
t\right) =\frac{1}{N}\frac{\dot{a}}{a}$.\ At this point we would like to
remark that we have assumed the comoving observer $u^{a}=\frac{1}{N}\delta
_{t}^{a}$, which $u^{a}u_{a}=-1$.

\subsection{MINISUPERSPACE LAGRANGIAN}

The field equations (\ref{me.06})-(\ref{me.08}) are of second-order in terms
of the parameters $\left\{ a,\phi\right\} $. Moreover, what it is
interesting is that the field equations can be derived under the variation
of the Lagrange function%
\begin{equation}
L\left( N,a,\dot{a},\phi,\dot{\phi}\right) =\frac{1}{N}\left( -3a\dot {a}%
^{2}+\frac{1}{2}a^{3}\dot{\phi}^{2}\right) -Na^{3}V\left( \phi\right) .
\label{me.09}
\end{equation}

More specifically, equations (\ref{me.07}), (\ref{me.08}) correspond to the
Euler-Lagrange equations of\ (\ref{me.09}) with respect to the variables $%
\left\{ a,\phi\right\} $, while variation with respect to the lapse function
$N$ provides the constraint equation (\ref{me.06}).

From (\ref{me.09}) someone can define the momentum $p_{a}=\frac{\partial L}{%
\partial\dot{a}}$~and~$p_{\phi}=\frac{\partial L}{\partial\dot{\phi}}$ and
write the Hamiltonian function%
\begin{equation}
\mathcal{H}\equiv N\left( -\frac{p_{a}^{2}}{3a}+\frac{p_{\phi}}{2a^{3}}%
+a^{3}V\left( \phi\right) \right)   \label{me.10}
\end{equation}
which from ~the constraint it follows that $\mathcal{H}=0$. At this point we
would like to remark that the field equations define the singular
Hamiltonian system, which is constrained by the conditions $p_{N}=0$ and (%
\ref{me.06}).

It is well known that there exists a unique relation between the symmetries
of dynamical systems defined by a kinetic energy and potential, equations
with the collineations which define the space where the motion occurs, i.e.
the kinetic metric. More specifically it is known that any generator of a
symmetry vector for the dynamical system has to be a symmetry also for the
geometry \cite{gnature}. For instance, the conservation law of momentum for
the free particle follows from the translation symmetry of the Euclidean
spacetime. The group of translations with the group of rotations form the
group of isometries or Killing vectors of the Euclidean space.

By definition a Killing vector in a Riemannian manifold is the generator of
the transformation which keeps invariant the length and the angles. On the
other hand, a Homothetic vector is the generator of the transformation which
keeps invariant the angles and rescales by a constant the length, whereas a
Conformal vector is called the generator of the transformation which
preserves the angles on the space \cite{sym5}. For autonomous Hamiltonian
systems the \textquotedblleft Energy\textquotedblright\ denotes the volume
in the phase space. For any isometry which leaves this volume invariant in
the phase space corresponds a conservation law which commutes with the
Hamiltonian. As far as the Homothetic vector concerned, the solutions can be
transformed under other solutions but with a rescaled \textquotedblleft
Energy\textquotedblright\ value. These two transformations relate objects
which are congruent, with the identical congruent to be provided by the
isometries. The situation is totally different under conformal
transformations. Indeed Hamiltonian systems are not invariant under
conformal transformations except if the \textquotedblleft
Energy\textquotedblright\ is zero, which means that the volume in the phase
space has dimensions zero. Moreover the volume continues to be zero under
conformal transformations and consequently conservation laws can be
constructed.

Mathematically, that is demonstrated as follows, consider $\mathcal{H}\left(
\mathbf{p,q}\right) =0$ to be the energy of an autonomous Hamiltonian system
and $I\left( \mathbf{p,q}\right) $ be a conservation law generated by a
conformal vector. Then it follows that there exists a function, $\omega$,
such that $D_{t}\left( I\right) =I_{,t}+\left\{ I,\mathcal{H}\right\} =\omega%
\mathcal{H}$; that is, $D_{t}\left( I\right) =0$, which means that $I$ is a
conservation law. These kinds of conservation laws are generated by nonlocal
symmetries, which are reduced to local when $\omega=const$ or $\omega=0$.

Because of the constraint equation we can say that the Energy of the
Mechanical analogue is zero and construct conservation laws by using the
conformal algebra of the minisuperspace. In particular, for every Conformal
vector field there corresponds a conservation law for the field equations,
for any function, $V\left( \phi\right) $. Moreover, because the
minisuperspace of (\ref{me.09}) has dimension two, it admits an
infinite-dimensional conformal algebra, that is, there exists an infinite
number of (nonlocal) conservation laws. Of course these conservation laws
are not in involution with each other, but they are with the Hamiltonian
applying the constraint equation,~$\mathcal{H}\left( \mathbf{p,q}\right) =0$%
. That concept is easily generalized and in the case of conformal Killing
tensors which define conservation laws polynomial in terms of the momentum $%
\mathbf{p}$ \cite{ns03,ck1}.

As far as the dynamical system of our study is concerned, the existence of a
nonlocal conservation law plus the constraint equation (\ref{me.09}) is
sufficient to prove the integrability. The generic algebraic solution is
presented below.

\subsection{ALGEBRAIC SOLUTION OF SCALAR FIELD COSMOLOGY}

The precise meaning of the solution of a system of differential equations
can be cast in several ways. Three of these are: (a) a set of explicit
functions describing the variation of the dependent variables with the
independent variable(s) (Closed-form solutions); (b) the existence of a
sufficient number of independent explicit\ first integrals and invariants;
(c) the existence of a sufficient number of explicit transformations which
permits the reduction of the system of differential equations to a system of
algebraic equations.

Hence, with the use of conservation laws generated by conformal vector
fields, it is possible to write the generic algebraic solution of the field
equations in scalar field cosmology. Indeed, the generic solution is given
in terms of an arbitrary function $F\left( \omega\right) $ \ as follows \cite%
{dimprd}%
\begin{equation}
\phi(\omega)=\pm\frac{\sqrt{6}}{6}\int\!\!\sqrt{F^{\prime}(\omega)}%
d\omega,~V(\omega)=\frac{1}{12}e^{-F(\omega)}\left(
1-F^{\prime}(\omega)\right)   \label{so.01}
\end{equation}
and%
\begin{equation}
\rho_{\phi}(\omega)=\frac{1}{12}e^{-F(\omega)}~,~P_{\phi}(\omega)=\frac{1}{12%
}e^{-F(\omega)}\left( 2F^{\prime}(\omega)-1\right) .   \label{so.03}
\end{equation}
where now the spatially flat FLRW spacetime is written as%
\begin{equation}
ds^{2}=-e^{F\left( \omega\right)
}d\omega^{2}+e^{\omega/3}(dx^{2}+dy^{2}+dz^{2}).   \label{SF.12}
\end{equation}

It is easy to see that $F\left( \omega\right) $ is related with the Hubble
functions as $F\left( \omega\right) =-2\ln\left( 3H\left( \omega\right)
\right) $. The latter solution is generic analytical solution for arbitrary
potential. The form of the potential fixes the EoS and provides an first
order-differential equation $p_{\phi}\left( \omega\right) =\Phi\left(
\rho_{\phi}\right) $ which can be reduced to an algebraic equation.
Otherwise, a specific function $F\left( \omega\right) $, provides always
(locally) a specific potential $V\left( \phi\left( \omega\right) \right) $.
\ Closed-form solutions where that scalar field admit a specific equation of
state parameter presented in \cite{dimprd,bar1}

\subsection{INFLATIONARY SLOW-ROLL PARAMETERS}

From the field equations (\ref{me.06}), (\ref{me.07}) we can observe that in
order the universe to be in an inflationary era, then the scalar field
potential dominates the kinetic term i.e., $\frac{\dot{\phi}^{2}}{2}<V(\phi)$%
, while the field equations can be written as
\begin{equation}
3H^{2}\simeq V\left( \phi\right) ~,~3\dot{\phi}\simeq V_{,\phi}
\label{me.11}
\end{equation}

Consequently, the potential slow-roll parameters (PSR), $\varepsilon
_{V}=\left( \frac{V_{,\phi}}{2V}\right) ^{2}~\,,~\eta_{V}=\frac{V_{,\phi
\phi}}{2V},~$have been introduced \cite{slp01} in order to study the
existence of the inflationary phase of the universe. Inflation occurs when $%
\varepsilon_{V}<<1$, while in order for the inflationary phase to last long
enough we require the second PSR parameter also to be small, $\eta_{V}<<1$.
That is a requirement for the flatness of the scalar field potential.

On the other hand, it is possible to express the inflation in terms of the
Hubble slow-roll parameters (HSR) which are defined similarly \cite{slpv}, $%
~\varepsilon_{H}=\left( \frac{H_{,\phi}}{H}\right) ^{2},~\eta_{H}=\frac{%
H_{,\phi\phi}}{H}.$ while the two different sets of slow-roll parameters are
related as follows~$\varepsilon_{V}\simeq\varepsilon_{H}~$and~$%
\eta_{V}\simeq\varepsilon_{H}+\eta_{H}.$

Because of the generic solution presented in the previous section we are
able to write the HSR parameters in terms of function $F\left( \omega\right)
$ and its derivatives as follows \cite{bar2}%
\begin{equation}
\varepsilon_{H}=3F^{\prime}~,~\eta_{H}=3\frac{\left( F^{\prime}\right)
^{2}-F^{\prime\prime}}{F^{\prime}}.   \label{me.12}
\end{equation}

The HSR can be used to define dimensionless observable parameters which are
related to the inflation, the spectral index for the density perturbations $%
n_{s}\equiv1-4\varepsilon_{H}+2\eta_{H},~$and the tensor to while the tensor
to scalar ratio is $r=10\varepsilon_{H}$. Moreover, the recent data analysis
by the Planck collaboration \cite{data2}, it was found that the value of $%
n_{s}$ with the error is $n_{s}=0.968\pm0.006,~$while the range of the
scalar spectral index is $n_{s}^{\prime}=-0.003\pm0.007$; while parameter $r$%
, has been found that it has an upper boundary, that is, $r<0.11$.

In \cite{bar2}, Barrow \&\ Paliathanasis, considered that the spectral index
$n_{s}$ and the scalar ration $r$ are related under with the relation~$%
n_{s}-1=f\left( r\right) ~$in which $f\left( r\right) $ is an analytic
functions. Moreover, with the use of (\ref{me.12}) the latter expression
defines a nonlinear function differential equation. Analytic solutions of
the latter equations which fit the inflationary data were determined for the
case in which $f\left( r\right) $ is constant, linear, or quadratic
function. This specific study is mainly focused on the symmetry of these
three master equations.

\section{MAXIMALLY SYMMETRIC EQUATIONS}

\label{sec3}

The three master equations of our consideration are
\begin{equation}
F^{\prime\prime}+\left( F^{\prime}\right) ^{2}-\frac{n_{0}}{3}F^{\prime}=0,
\label{ms.01}
\end{equation}%
\begin{equation}
F^{\prime\prime}+\left( 1-n_{1}\right) \left( F^{\prime}\right) ^{2}-\frac{%
n_{0}}{3}F^{\prime}=0,   \label{ms.02}
\end{equation}
and%
\begin{equation}
F^{\prime\prime}+3n_{2}\left( F^{\prime}\right) ^{3}+\left( 1-n_{1}\right)
\left( F^{\prime}\right) ^{2}-\frac{n_{0}}{3}F^{\prime}=0.   \label{ms.03}
\end{equation}

We apply Lie's theory \cite{Bluman} and we calculate the Lie point
symmetries for the three master second-order differential equations.

For equation (\ref{ms.01}) the Lie point symmetries are%
\begin{equation}
X_{1}=\partial _{\omega }~,~X_{2}=\partial _{F}~,~X_{3}=e^{F-\frac{n_{0}}{3}%
\omega }\partial _{\omega }~,~X_{4}=e^{\frac{n_{0}}{3}-F}\partial _{F}~,
\end{equation}%
\begin{equation}
X_{5}=e^{-F}\partial _{F}~,~X_{6}=e^{-\frac{n_{0}}{3}\omega }\partial
_{\omega },~X_{7}=e^{F}\left( 3\partial _{\omega }+n_{0}\partial _{F}\right)
~,
\end{equation}%
\[
X_{8}=e^{\frac{n_{0}}{3}\omega }\left( 3\partial _{\omega }+n_{0}\partial
_{F}\right)
\]

On the other hand, the Lie point symmetries of equation (\ref{ms.02}) are
calculated to be%
\begin{equation}
Y_{1}=\partial _{\omega }~,~Y_{2}=\partial _{F}~,~Y_{3}=e^{\left(
n_{1}-1\right) F-\frac{n_{0}}{3}\omega }\partial _{\omega }~,~Y_{4}=e^{\frac{%
n_{0}}{3}-\left( n_{1}-1\right) F}\partial _{F}~,
\end{equation}%
\begin{equation}
Y_{5}=e^{-\left( n_{1}-1\right) F}\partial _{F}~,~Y_{6}=e^{-\frac{n_{0}}{3}%
\omega }\partial _{\omega },~Y_{7}=e^{\left( n_{1}-1\right) F}\left(
3\partial _{\omega }+n_{0}\partial _{F}\right) ~,
\end{equation}%
\[
Y_{8}=e^{\frac{n_{0}}{3}\omega }\left( 3\partial _{\omega }+n_{0}\partial
_{F}\right)
\]

Finally, for equation (\ref{ms.03}) the Lie point symmetries are%
\begin{equation}
Z_{1}=\partial _{\omega }~,~Z_{2}=\exp \left( -\frac{n_{0}}{3}-\frac{\left(
n_{1}-1+A\left( n_{0},n_{1},n_{2}\right) \right) }{2}F\right) \partial
_{\omega }~,~
\end{equation}%
\begin{equation}
Z_{3}=\exp \left( -\frac{n_{0}}{3}+\frac{\left( n_{1}-1+A\left(
n_{0},n_{1},n_{2}\right) \right) }{2}F\right) \partial _{\omega
}~,~Z_{4}=\left( 3-3n_{1}\right) \partial _{\omega }+2n_{0}\partial _{F}~,~
\end{equation}%
\begin{equation}
Z_{5}=\exp \left( -\frac{n_{0}}{3}-\frac{\left( n_{1}-1+A\left(
n_{0},n_{1},n_{2}\right) \right) }{2}F\right) \partial _{F} \\
Z_{6}=\exp \left( -\frac{n_{0}}{3}+\frac{\left( n_{1}-1+A\left(
n_{0},n_{1},n_{2}\right) \right) }{2}F\right) \partial _{F}
\end{equation}%
\begin{equation}
Z_{7}=2n_{0}\cosh \left( A\left( n_{0},n_{1},n_{2}\right) F\right) \partial
_{F}-3\left( \left( n_{1}-1\right) \cosh \left( A\left(
n_{0},n_{1},n_{2}\right) F\right) -\sinh \left( A\left(
n_{0},n_{1},n_{2}\right) F\right) \right) \partial _{\omega }
\end{equation}%
\[
Z_{8}=2n_{0}\sinh \left( A\left( n_{0},n_{1},n_{2}\right) F\right) \partial
_{F}-3\left( \left( n_{1}-1\right) \sinh \left( A\left(
n_{0},n_{1},n_{2}\right) F\right) -\cosh \left( A\left(
n_{0},n_{1},n_{2}\right) F\right) \right) \partial _{\omega }
\]%
where constant $A\left( n_{0},n_{1},n_{2}\right) $ is defined as $A\left(
n_{0},n_{1},n_{2}\right) =\sqrt{\left( n_{1}-1\right) ^{2}+4n_{0}n_{2}}$.

We note with surprise that the three master equations (\ref{ms.01})-(\ref%
{ms.03}) admit eight Lie point symmetries and are maximally symmetric. It is
well-known that all the second-order differential equations which are
invariant under the action of the $SL\left( 3,R\right) $ group, are
equivalent under point transformations \cite{lie1}. Hence, there exist point
transformations which transform any solution of the master equations to
another solution.

For instance, consider equation (\ref{ms.01}) with $n_{0}=0$, then, under
the point transformation~$F\left( \omega\right) \rightarrow\left(
1-n_{1}\right) \bar{F}\left( \omega\right) ,$equation (\ref{ms.01}) is
written as%
\begin{equation}
\bar{F}^{\prime\prime}+\left( 1-n_{1}\right) \left( \bar{F}^{\prime }\right)
=0
\end{equation}
which is nothing else than equation (\ref{ms.02}). However, at the same time
the line element of the spacetime (\ref{SF.12}) transformed as follows
\begin{equation}
ds^{2}=-\left( e^{-\bar{F}\left( \omega\right) }\right) ^{\left(
1-n_{1}\right) }d\omega^{2}+e^{\omega/3}(dx^{2}+dy^{2}+dz^{2}).
\label{in.61}
\end{equation}

Consider now the classical Newtonian analogue of a free particle and an
observer whose measuring instruments for time and distance are not linear.
By using the measured data of the observer we reach in the conclusion that
it is not a free particle. On the other hand, in the classical system of the
harmonic oscillator an observer with nonlinear measuring instruments can
conclude that the system observed is that of a free particle, or that of the
damped oscillator or another system. From the different observations,
various models can be constructed. However, all these different models
describe the same classical system and the master equations are invariant
under the same group of point transformations but in a different
parametrization.

In the master equations that we studied there is neither position nor time
variables: the independent variable is the scale factor $\omega=6\ln a$, and
the Hubble function is the dependent variable, $H\left( a\right) $.
Therefore, we can say that at the level of the first-order approximation for
the spectral indices, various representations of the variables $\left\{
a,H\left( a\right) \right\} $ provide different observable values for the
spectral indices. Hence we showed that the method of Lie symmetries can be
used to determine new solutions from others.

\section{CONCLUSIONS}

\label{sec4}

In this article we discussed the application of symmetries for differential
equations in cosmology with emphasis in the inflationary era. Because of the
complexity of the nonlinear gravitational field equations Lie symmetries can
play an important role in the study of the dynamical evolution of the
cosmological models as also in the determination of exact and analytic
solutions.

In the case of inflationary models, the novelty of the Lie's theory is that
point transformations can be applied to determine new solutions from
existing ones, while to transform known solutions from one to the other. The
latter is possible by applying the Lie theorem for maximally symmetric
second-order differential equations.

\end{document}